\documentclass[aps,twocolumn,showpacs]{revtex4}

\begin{document}
\title{Structural physical approximations of unphysical maps and
generalized quantum measurements}
\author{Jarom\'{\i}r Fiur\'{a}\v{s}ek}
\affiliation{Department of Optics, Palack\'{y} University, 17. listopadu
50, 77200 Olomouc, Czech Republic}

\begin{abstract}
We investigate properties of the structural physical approximation
(SPA) of the partial transposition map recently introduced by
Horodecki and Ekert [quant-ph/0111064].
We focus on the case of two-qubit states and show that in this case
the map has the structure of a generalized quantum measurement
followed by preparation of a suitable output state.
We also introduce  SPA
for map that transforms two copies of density matrix of a single qubit
onto a square of that matrix. We prove that also this map is essentially
a generalized quantum measurement.
\end{abstract}

\maketitle

Completely positive (CP) maps represent the most general transformations
of quantum states \cite{Kraus83}.
The fact that quantum dynamics is described by linear CP maps on density
matrices can be derived from the static properties of quantum mechanics
and the no-signaling condition which states that superluminal
communication is excluded \cite{Simon01}. The formalism of CP maps finds
a wide variety of applications in the rapidly growing field of Quantum
Information Theory. CP maps are capable to describe
an arbitrary quantum transmission channel \cite{Schumacher96}.
Construction of devices that optimally approximate some
unphysical transformations, such as quantum cloners \cite{Buzek96}
or universal-NOT gate \cite{Gisin99,Buzek99},
can be formulated in a unified framework as a determination of an optimal
CP map \cite{Fiurasek01}. Properties of probabilistic and
deterministic transformations between sets of pure states
can be conveniently analyzed with the help of the formalism
of CP maps \cite{Chefles01}.

The map $\cal{E}$ is called positive if for any $\rho\geq0$ it holds that
${\cal{E}}(\rho)\geq 0$. The map is completely positive iff the induced map
${\cal{E}}\otimes {\cal{I}}_h$ is positive for all $h$, where
${\cal{I}}_h$ is an identity map on an auxiliary $h$ dimensional Hilbert
space. The positive maps that are not completely positive play central role
in detecting quantum entanglement. It was shown by Horodeckis
\cite{Horodecki96} that a bipartite state $\rho_{AB}$ is separable if
$[{\cal{I}}_A\otimes{{\cal{M}}}_B](\rho_{AB})$
is a positive semidefinite operator
for any positive map $\cal{M}$  which is not a CP-map. Particularly
important is the partial transposition map ${\cal{P}}={\cal{I}}_A\otimes
{\cal{T}}_B$, whose ability to detect the entanglement was first pointed
out by Peres \cite{Peres96}. Since  $\cal{P}$ is not a CP map,
it seems to be impossible to implement this map physically in a lab.
Recently, however, an ingenious
way how to circumvent this obstacle was suggested by Horodecki and Ekert
\cite{Horodecki01a,Horodecki01b}, who introduced the so-called
structural physical approximation (SPA) of the unphysical map $\cal{P}$.
The idea is to form a mixture of the map $\cal{P}$
with a CP map $\cal{O}$ that transforms all quantum states onto maximally mixed
state, ${\cal{O}}(\rho)= \openone/d$, where $\openone$ is a unit operator
and $\rho$ is an arbitrary operator acting on $d$-dimensional Hilbert space.
In the case of the partial
transposition map acting on state of two qudits,
the optimal SPA is given by \cite{Horodecki01b}
\begin{equation}
{\cal{P}}_{S}= \frac{d^3}{d^3+1}
{\cal{O}}_A\otimes {\cal{O}}_B +\frac{1}{d^3+1}{\cal{I}}_A\otimes{\cal{T}}_B.
\label{PS}
\end{equation}

Horodecki and Ekert \cite{Horodecki01b} suggested,
that the physical trace-preserving CP map ${\cal{P}}_S$ can be utilized
to experimentally directly detect the entanglement
without the necessity to carry out a full tomographic reconstruction
of the bipartite state $\rho_{AB}$ whose entanglement is to be determined.
Briefly, the proposed protocol goes as follows. One applies the map
(\ref{PS}) to many
copies of the bipartite state and then measures the spectrum of the
transformed state. This requires determination of no more than $d^2-1$
parameters, in contrast to $d^4-1$ parameters needed for a full
reconstruction of the state $\rho_{AB}$. For example, if one possesses $N$ copies
of the state $\rho$, then one can measure the quantity
$\mu_N=\langle \rho^N \rangle$ which is equal to sum of $N$-th powers of
eigenvalues of $\rho$. If  the moments $\mu_N$, $N=2,\ldots, d^2$,
are known, then the eigenvalues can be calculated simply by solving
a system of algebraic equations. It this context, an interesting question is whether one can
find SPA map that would approximate the transformation
\begin{equation}
\rho^{\otimes N} \rightarrow \rho^N .
\label{RHONPOWER}
\end{equation}
Horodecki conjectured that such SPA trace preserving map does not exist
\cite{Horodecki01a}.

In this paper we shall analyze the structure of the map ${\cal{P}}_S$ acting on
two-qubit state, $d=2$. We shall show that ${\cal{P}}_S$ can be accomplished by
a measurement based scenario: certain generalized measurement is carried
out on  the two-qubit state $\rho_{AB}$ and the output state is prepared
according to the result of this measurement. We shall discuss
consequences of this fact for the direct entanglement detection
protocol of Horodecki and Ekert.  In the second part of the paper
 we show that  SPA of the transformation (\ref{RHONPOWER}) can be
constructed if we allow this map to be trace decreasing. We shall provide
an explicit formula for $d=2$ and $N=2$ and  it will become clear that
an extension to higher dimensions and/or higher powers $N$ is quite
straightforward.

Our analysis of the map (\ref{PS}) is based on the well known correspondence
between linear completely positive maps and positive semidefinite operators
(see, e.g., \cite{Fiurasek01}). Consider a map $\cal{E}$ which
transforms operators acting on input
Hilbert space $\cal{H}$ onto operators acting on output Hilbert space
$\cal{K}$. The map $\cal{E}$ is uniquely specified by a positive
semidefinite operator $E$ defined as follows,
\begin{equation}
E =[{\cal{E}} \otimes {\cal{I}}](\Phi),
\label{EDEF}
\end{equation}
where $\cal{I}$ is an identity map, $\Phi=|\phi\rangle\langle\phi|$ and
\begin{equation}
|\phi\rangle=\sum_{j=1}^{{\rm dim}{\cal{H}}} |j\rangle|j\rangle
\label{PHIDEF}
\end{equation}
is a maximally entangled state on ${\cal{H}}^{\otimes 2}$.
This correspondence is not only mathematical, the state $E$ can be
physically prepared in the lab from the entangled state $\Phi$ by
applying the map $\cal{E}$ to one part of the state $\Phi$. This
allows one, for instance, to transform the problem of tomographic
reconstruction of an unknown quantum channel $\cal{E}$ onto
problem of reconstruction of an unknown quantum state $E$
\cite{Fischer01,DAriano01}, where
a variety of well established tomographic techniques may be applied.
Moreover, notions established for quantum states can be
straightforwardly extended to CP maps. One can define fidelity of two
CP maps \cite{Raginsky01}, and consider entanglement \cite{Zanardi01},
storage, compression and purification of the CP maps \cite{Dur01}.

In our case both the input and output Hilbert spaces are spaces
of two qubits and the operator $P_S$ which represents the map
\begin{equation}
{\cal{P}}_{S}= \frac{8}{9} {\cal{O}}\otimes {\cal{O}}
+\frac{1}{9} {\cal{I}}\otimes {\cal{T}}
\label{PSQUBIT}
\end{equation}
can be obtained if we insert (\ref{PSQUBIT}) into Eq. (\ref{EDEF}).
We show that the map ${\cal{P}}_S$ is {\em input-output separable}, i.e.,
that the operator $P_S$ can be written as a following convex sum,
\begin{equation}
P_S=\sum_{j} \Pi_j^T \otimes \rho_j,
\label{PSSEP}
\end{equation}
where $T$ stands for the transposition and
 $\Pi_j$ and $\rho_j$ are positive semidefinite Hermitian
operators  on  input and output Hilbert spaces $\cal{H}$ and
$\cal{K}$, respectively,
\begin{equation}
\Pi_j \geq 0,  \qquad \rho_j\geq 0.
\label{PIRHO}
\end{equation}
Moreover, $\rho_j$ represent density matrices, hence their traces are
normalized,
\begin{equation}
Tr[\rho_j]=1.
\label{TRACE}
\end{equation}
Furthermore, the operators $\Pi_j$ form elements of a positive operator
valued measure (POVM) and their sum is a unit operator,
\begin{equation}
\sum_j \Pi_j=\openone_{\cal{H}}.
\label{SUMPI}
\end{equation}
To prove this we recall that the map ${\cal{P}}_S$ is trace preserving.
In terms of the operator $P_S$ this condition reads \cite{Fiurasek01}
\begin{equation}
Tr_{\cal{K}}[P_S]=\openone_{\cal{H}}.
\label{TRACEPRES}
\end{equation}
On inserting the formula (\ref{PSSEP}) into Eq. (\ref{TRACEPRES})
and making use of the normalization of $\rho_j$ we
immediately arrive at (\ref{SUMPI}).

Let us now see what are the implications of the separability of the
operator $P_S$. The relation between input and output density matrices
$\rho_{\rm out}= {\cal{P}}_S(\rho_{\rm in})$ can be written with the
help of $P_S$ as follows \cite{Fiurasek01},
\begin{equation}
\rho_{\rm out}=Tr_{\cal{H}}[ P_S \rho_{\rm in}^T \otimes\openone_{\cal{K}}]
\label{RHOOUT}
\end{equation}
where $Tr_{\cal{H}}$ stands for the partial trace over the input Hilbert
space. On inserting the convex sum (\ref{PSSEP}) into Eq.
(\ref{RHOOUT}), we obtain
\begin{equation}
\rho_{\rm out}=\sum_j \rho_j Tr[\rho_{\rm in} \Pi_j].
\label{RHOOUTSEP}
\end{equation}
The interpretation of this expression is rather straightforward, the CP
map ${\cal{P}}_S$  is essentially a generalized measurement described by the
POVM elements $\Pi_j$ and carried out on the input two-qubit state
$\rho_{\rm in}$. If the $j$-th outcome is detected, then the
output state is prepared in a state $\rho_j$. Notice that $\Pi_j$ is
detected with the probability $p_j=Tr[\rho_{\rm in}\Pi_j]$.

We now prove that the operator $P_S$  is input-output separable.
We do so by explicitly rewriting $P_S$ in the form of convex sum of
tensor products of positive semidefinite operators in input and output
Hilbert spaces. After a somewhat lengthy but straightforward algebra one
arrives at the following decomposition,
 \begin{eqnarray}
P_S&=&\frac{1}{9}\left[W_{\,0010}^{0010}+W_{\,0111}^{0111}+W_{\,0001}^{0100}
+W_{\,0011}^{0110}\right. \nonumber \\[2mm]
&& +W_{\,1110}^{1011}
+W_{\,1001}^{1100} +S_{00}^{00}+S_{11}^{11}+S_{10}^{10}+S_{01}^{01}
\nonumber \\[2mm]
&&\left.+2\left(S_{11}^{00}+S_{00}^{11}+S_{10}^{01}+S_{01}^{10}\right)\right],
\label{PSEXPLICIT}
\end{eqnarray}
where
\begin{equation}
   W_{ijkl}^{abcd}=S_{ij}^{ab}+S_{ij}^{cd}+S_{kl}^{ab}+S_{kl}^{cd}
+F_{ijkl}^{abcd}+F_{klij}^{cdab},
\label{W}
\end{equation}
and
\begin{eqnarray}
&S_{ij}^{ab}=|ij\rangle\langle ij|\otimes |ab\rangle\langle ab|,&
\nonumber \\[2mm]
&F_{ijkl}^{abcd}=|ij\rangle\langle{kl}|\otimes |ab\rangle\langle cd|.&
\label{S}
\end{eqnarray}
Here the letters $i,j,k,l$ label states in $\cal{H}$ while $a,b,c,d$
denote states in $\cal{K}$. We still have to prove that the operators
$W_{ijkl}^{abcd}$ are separable. To do so, we notice that all operators
$W$ have the same structure.
For the sake of notational simplicity, we make the identification
\begin{eqnarray*}
|ij\rangle=|\uparrow\rangle_1,  \qquad
|kl\rangle=|\downarrow\rangle_1, \\
|ab\rangle=|\uparrow\rangle_2, \qquad
|cd\rangle=|\downarrow\rangle_2.
\end{eqnarray*}
Thus each operator $W$ represents a state of two qubits.
In the basis $|\uparrow\uparrow\rangle_{12}$,
$|\uparrow\downarrow\rangle_{12}$,
$|\downarrow\uparrow\rangle_{12}$, $|\downarrow\downarrow\rangle_{12}$,
the matrix representation of the operator $W$ reads
\begin{equation}
W=\left(
\begin{array}{cccc}
1 & 0 & 0 & 1 \\
0 & 1 & 0 & 0 \\
0 & 0 & 1 & 0 \\
1 & 0 & 0 & 1
\end{array} \right).
\label{WMATRIX}
\end{equation}
The separability of the state (\ref{WMATRIX})
can be easily proven with the help of
partial transposition criterion, because the condition
$[{\cal{I}}_1\otimes {\cal{T}}_2](W)\geq 0$
is necessary and sufficient for separability of the state $W$.
We can even provide an explicit decomposition of $W$ into
convex sum of tensor products,
\begin{equation}
W=\frac{1}{4}\sum_{j=1}^4 |\phi_j\rangle\langle\phi_j|
\otimes |\psi_j\rangle\langle\psi_j|,
\label{WSEP}
\end{equation}
where the states $|\phi_j\rangle \in \cal{H}$ and
$|\psi_j\rangle \in \cal{K}$ are given by
\begin{eqnarray*}
|\phi_j\rangle &=&|\uparrow\rangle+e^{i\frac{\pi}{2}j}|\downarrow\rangle,
\\
|\psi_j\rangle &=&|\uparrow\rangle+e^{-i\frac{\pi}{2}j}|\downarrow\rangle.
\end{eqnarray*}
This concludes our proof of the input-output separability of the map
${\cal{P}}_S$.

We now discuss the implications of our results for the entanglement
detection scheme proposed by Horodecki and Ekert. Since the CP map
${\cal{P}}_S$ is essentially generalized quantum measurement, the quantum
information contained in the two-qubit state $\rho_{\rm in}$
is converted into a classical information represented by the results of
the generalized measurement. Moreover, the preparation of the
output state $\rho_j$ can be repeated as many times as we wish.
The situation here is very similar to the optimal universal-NOT gate,
where also the measurement based scenario is the best one \cite{Buzek99}
and an arbitrary number of the flipped spins can be prepared at the
output, all with the same fidelity.

The measurement-based scenario for the CP map ${\cal{P}}_S$ also indicates
that any subsequent measurement on the output states
$\rho_{\rm out}={\cal{P}}_S(\rho_{\rm in})$ can not reveal any more
information in addition to that obtained from the POVM measurement
(\ref{SUMPI}).
Suppose that we would like to first apply the map (\ref{PSQUBIT})
to $N$ copies of the two-qubit state $\rho_{\rm in}$ and then carry
out some collective measurement on the output state $\rho_{\rm
out}^{\otimes N}$. We can accomplish this if we perform the
measurement (\ref{SUMPI}) on each copy of the input state, then we prepare the
output states $\rho_j$ according to the measurement results and finally
carry out the collective measurement. However, from the results of the
measurements (\ref{SUMPI}) on the input states we actually exactly know
the particular output state. Hence we need not to prepare the output
states  $\rho_j$ and
carry out the collective measurements. Instead, we can simulate the measurement
process on a classical computer or even simply  calculate mean value of
any operator for the particular output state. This
implies that the necessity of storage of $N$ copies of the output
quantum states is avoided.

Furthermore, the generalized measurement (\ref{SUMPI}) is
tomographically complete.
This can be easily proved as follows. If we know exactly all the
probabilities $p_j$ then we can reconstruct the output state
$\rho_{\rm out}$. From this output state we can then obtain the input
state $\rho_{\rm in}$ simply by inverting the linear relation
established by the CP map ${\cal{P}}_S$. We first subtract a multiple of
identity operator and then we apply a partial transposition map (which
is its own inverse),
\begin{equation}
\rho_{\rm in}={\cal{P}}(9\,\rho_{\rm out}-2\,\openone).
\label{RHOIN}
\end{equation}
In this way we recover the input state $\rho_{\rm in}$.
These considerations seem to indicate that a tomographically
complete measurement is  ``hidden'' in the CP map ${\cal{P}}_S$
used in the scenario for direct detection of the entanglement suggested
by Horodecki and Ekert \cite{Horodecki01b}.
This tomographically complete measurement
also seems to provide more information than any subsequent (possibly
collective) measurement on the output states. Nevertheless, the claim of
Horodecki and Ekert that their protocol involves determination of a
reduced number of parameters in comparison to full tomographic
measurement remains valid. However, the noise introduced by the part of
the map ${\cal{P}}_S$ proportional to ${\cal{O}}\otimes \cal{O}$ reduces the
amount  of information that can be extracted from the (collective)
measurements on the output states so it seems that these measurements
cannot be more efficient than the POVM (\ref{SUMPI}).

In the rest of the paper, we shall construct an structural physical
approximation  for the transformation (\ref{RHONPOWER}).
A key observation is that the
transformation $\rho^{\otimes N}\rightarrow \rho^N$ is in certain sense
linear. Indeed, the matrix elements of $\rho^N$ are linear combinations
of the matrix elements of the operator $\rho^{\otimes N}$.
For the sake of simplicity, we shall focus  on making
square of a density matrix of a single qubit, $d=2$ and $N=2$.
Since the map
\begin{equation}
{\cal{R}}(\rho\otimes\rho)=\rho^2
\label{RMAP}
\end{equation}
is linear, it can be represented by an operator
$R$ according to Eq. (\ref{EDEF}).
We note that the operator $R$ is not positive semidefinite, because
(\ref{RMAP}) is not a CP map. However, we can define the
operation $\cal{R}$ in such way that it preserves the hermiticity property.
This is equivalent to the fact that the corresponding
operator $R$ is Hermitian,
\begin{eqnarray}
R&=&\left[|00\rangle\langle 00|+\frac{1}{2}\left(|01\rangle\langle
10|+|10\rangle\langle 01|\right)\right] \otimes|0\rangle\langle 0|
\nonumber \\
&&+\left[|11\rangle\langle 11|+\frac{1}{2}\left(|01\rangle\langle
10|+|10\rangle\langle 01|\right)\right] \otimes|1\rangle\langle 1|
\nonumber \\
&&+\frac{1}{\sqrt{2}}\biggl[|00\rangle\langle +|+
|+\rangle\langle 11| \biggr]\otimes|0\rangle\langle 1|
\nonumber \\
&&+\frac{1}{\sqrt{2}}\biggl[|+\rangle\langle 00|+
|11\rangle\langle +| \biggr]\otimes|1\rangle\langle 0|,
\end{eqnarray}
\label{R}
where$|+\rangle=(|01\rangle +|10\rangle)/\sqrt{2}$.

Now, it was shown by Horodecki \cite{Horodecki01a}, that there exists
structural physical approximation for {\em any} map that preserves
hermiticity property. The explicit construction involves in our case
a mixture of
the operation ${\cal{R}}$ and an operation $\bar{\cal{O}}$ that maps
any two-qubit density matrix onto maximally mixed state of a single qubit,
$\bar{{\cal{O}}}(\rho)=\openone/2$. The weight of the operation
$\bar{\cal{O}}$
in the mixture is equal to the lowest eigenvalue of the operator
$R$. In this way we define SPA operation that optimally approximates the
unphysical transformation $\cal{R}$,
\begin{equation}
{\cal{R}}_S=\frac{1}{2}(\bar{\cal{O}}+{\cal{R}}).
\label{RSMAP}
\end{equation}
The operator $R_S$ representing  CP-map ${\cal{R}}_S$
is given by
\vspace*{-2mm}
\begin{eqnarray}
R_S&=&\frac{1}{2}\left[\frac{3}{2}|00\rangle\langle 00|+|+\rangle\langle +|
+\frac{1}{2}|11\rangle\langle 11|\right] \otimes|0\rangle\langle 0|
\nonumber \\
&&+\frac{1}{2}\left[\frac{3}{2}|11\rangle\langle 11|+|+\rangle\langle+|
+\frac{1}{2}|00\rangle\langle 00|\right] \otimes|1\rangle\langle 1|
\nonumber \\
&&+\frac{1}{2\sqrt{2}}\biggl[|00\rangle\langle +|+
|+\rangle\langle 11| \biggr]\otimes|0\rangle\langle 1|
\nonumber \\
&&+\frac{1}{2\sqrt{2}}\biggl[|+\rangle\langle 00|+
|11\rangle\langle +| \biggr]\otimes|1\rangle\langle 0|.
\label{RS}
\end{eqnarray}
The CP-map ${\cal{R}}_S$ is trace decreasing, which is expressed by the
inequality $Tr_{\cal{K}}[R_S]< \openone$. Thus the map (\ref{RSMAP})
can be applied only probabilistically
(see, e.g., Appendix of Ref. \cite{Horodecki01a}),
and the probability of success is given by
\begin{equation}
P=\frac{1}{2}[1+Tr(\rho^2)].
\label{P}
\end{equation}

Since the matrix $\rho\otimes\rho$ is symmetric, the input Hilbert space
for the transformation ${\cal{R}}_S$ is the symmetric
subspace of the Hilbert space of two qubits which has dimension $3$.
The output Hilbert space is, of course, the Hilbert space of a single
qubit. We can thus easily check the input-output separability of the
CP map ${\cal{R}}_S$, because the Peres-Horodecki criterion is necessary and
sufficient in this case \cite{Horodecki96,Peres96}.
We find that the operator $R_S$ has positive
partial transpose hence is separable. We can write
 \begin{equation}
 R_S= \sum_{j} \bar{\Pi}_j^T \otimes \bar{\rho}_j
 \label{RSSEP}
 \end{equation}
and the map ${\cal{R}}_S$ again corresponds to some generalized measurement.
However, the map ${\cal{R}}_S$ is trace-decreasing, which means that
$\sum_j \bar{\Pi}_j < \openone.$
The measurement defined by POVM elements $\bar{\Pi}_j$
is incomplete and must be completed by adding one more element
$\bar{\Pi}_0=\openone-\sum_j\bar{\Pi}_j$. The map (\ref{RSMAP}) can be
implemented as follows. One carries out the completed measurement
$\bar{\Pi}_j$ on the input two-qubit state. If $\bar{\Pi}_0$
is detected, then the procedure failed. If any other measurement
outcome is observed, then the corresponding state $\bar{\rho}_j$
is prepared at the output.

In summary, we have studied properties of the structural physical
approximation of the partial transposition map on two qubits.
We have shown that the SPA map is equivalent to certain generalized
quantum measurement. This has consequences for the scenario for
direct detection of entanglement proposed by Horodecki and Ekert.
In their scheme, collective measurements on several copies of the output
states have to be performed. However, these measurements cannot be more
efficient than the POVM which is associated with the CP map
${\cal{P}}_S$. Our discussion was restricted to the case of two qubits and it
would be interesting to see whether the CP map (\ref{PS}) is equivalent to
generalized measurement also for higher dimensions. This is an open
question which certainly deserves further investigation.

We have also introduced an
SPA for the map that makes $\rho^2$ from two copies of the density
matrix of a single qubit $\rho$. We have shown that this SPA map is
trace-decreasing and is also equivalent to some generalized quantum
measurement. We expect that the construction of the map ${\cal{R}}_S$
can be quite straightforwardly extended to higher dimensions and/or powers,
which will be the subject of further investigations.

I would like to thank  R. Filip and Z. Hradil for stimulating
discussions. This work was supported by Grant No LN00A015 and
Research Project CEZ: J14/98 of the Czech Ministry of Education.

\end{document}